# Enhancing Dose Selection in Phase I Cancer Trials: Extending the Bayesian Logistic Regression Model with Non-DLT Adverse Events Integration

## Authors

Andrea Nizzardo[1], Luca Genetti[1], Marco Pergher[1]

[1]Pharmacometrics Team, Clinical Development and Translational Medicine, Evotec, Levi-Montalcini Campus, 4 A. Fleming Street, Verona (Italy), 35135

## Abstract

This work introduces the Burdened Bayesian Logistic Regression Model (BBLRM), an enhancement of the Bayesian Logistic Regression Model (BLRM) for dose-finding in phase I oncology trials. The BLRM determines the maximum tolerated dose (MTD) based on dose-limiting toxicities (DLTs). However, clinicians often perceive model-based designs like BLRM as complex and less conservative than rule-based designs, such as the widely used 3+3 method. To address these concerns, BBLRM incorporates non-DLT adverse events (nDLTAEs), which, although not severe enough to be DLTs, indicate potential toxicity risks at higher doses.

BBLRM introduces an additional parameter δ to account for nDLTAEs, adjusting toxicity probability estimates to make dose escalation more conservative while maintaining accurate MTD allocation. This parameter, generated basing on the proportion of patients experiencing nDLTAEs, is tuned to balance conservatism with model performance, reducing the risk of selecting overly toxic doses. Additionally, involving clinicians in identifying nDLTAEs enhances their engagement in the dose-finding process.

A simulation study compares BBLRM with two other BLRM methods and a two-stage Continual Reassessment Method (CRM) incorporating nDLTAEs. Results show that BBLRM reduces the proportion of toxic doses selected as MTD without compromising the accuracy in MTD identification. These findings suggest that integrating nDLTAEs can improve the safety and acceptance of model-based designs in phase I oncology trials.

## 1. Introduction

One primary objective of phase I oncology dose-finding trials is to identify the maximum tolerated dose (MTD) of the drug candidate to select the dose level(s) to be investigated in subsequent phases of development. In most cases, the data used to determine the MTD are a binary scale based on the presence or absence of dose-limiting toxicity (DLT) within a certain period/cohorts of patients enrolled into provisional dose levels, usually starting from the lowest or second lowest one, in a sequential and adaptive manner. Upon observing the DLT data from the latest cohort of patients, a recommendation is made for the dose level to be assigned to the next cohort of patients based on a specific dose-finding design. This process is repeated until the total sample size is exhausted or certain prespecified early stopping rules are met.

Different approaches have been developed and reported in the literature to put into practice the dose escalation procedure. These can be mainly divided into two categories: algorithm-rule-based and model-based designs. Moreover, there is another class of methods, the model-assisted designs, a hybrid approach that combines both the clear predetermined algorithm to guide the dose escalation in rule-based designs and the underlying statistical models adopted in model-based designs.

Algorithm-rule-based approaches consist of deterministic rules that are systematically applied during the trial. The decisions for the dose escalation process are based on the number of DLTs observed only in the current cohort, without considering all the information collected up to that moment. The most popular and widely adopted rule-based approach is the 3+3 method, introduced by Carter in 1987[1].

Model-assisted designs use statistical models (e.g., the binomial model) to derive the design for efficient decision making; however, like the algorithm-based design, the dose escalation and de-escalation rules must be predetermined before the onset of the trial, and they can be easily implemented as in the algorithm-based designs. The most popular approach of this type is the Bayesian optimal interval (BOIN[2]) design, which stands out compared to other methods in this class in terms of performance and flexibility, especially when compared to algorithm-based designs[3].

Model-based designs, instead, allocate patients to a dose level using a targeted toxicity rate and a statistical model that describes the dose-toxicity relationship between the dose levels; when a new cohort of patients is initiated to the trial, the model is updated using all available information on all previously enrolled patients and the dose for the new cohort is agreed using the model-suggested dose as a guideline. In 1990, O'Quigley J. et al[4] proposed the first model-based design, the Continual reassessment method (CRM), a one-parameter regression model to estimate the toxicity probability for a DLT event. Several developments from the original CRM have been proposed; in particular, Neuenschwander et al. in 2008[5] introduced the Bayesian logistic regression model (BLRM), a two-parameter model to specify better and thus estimate the dose-toxicity curve, considering also an escalation rule to control the overdosing probability, introduced originally in 1998 by Babb et al.[6] for the CRM model.

Alongside these new developments, several concerns have been raised about the quality of the operating characteristics of the 3+3 design, compared to the model-based designs[7,8,9]. Statistical simulations have demonstrated that the 3+3 design identifies the MTD in as few as 30% of trials[2]. Furthermore, some argue that this method for dose escalation may result in a high proportion of patients being treated at subtherapeutic doses[10].

From this perspective, model-assisted designs outperform rule-based designs. However, due to their simplicity compared to model-based approaches, they do not allow for the development of more sophisticated extensions that incorporate additional covariate information, drug combinations[11] (potentially with drug-drug interactions), or other types of data, such as pharmacokinetics[12,13,14] or animal data[15].

The 3+3 dose escalation method remains the most widely used approach in phase I trials[16,17,18,19]. Clinicians are often reluctant to adopt model-based designs for several reasons. They argue that such designs are less efficient, potentially prolonging trial duration and exposing more patients to toxic doses. Safety concerns also play a role, as the 3+3 method is viewed as more conservative and easier to implement. Additionally, clinicians value the ability to overrule dose escalation decisions, something they feel model-based approaches may limit. Finally, model-based designs are often perceived as black-box approaches, making their parameters harder to interpret during trial planning.

To overcome some of these concerns, extensions of the original model-based designs have been proposed by incorporating decision-stopping rules to reduce as much as possible the number of patients assigned to dose levels that are below (underdosing) or above (overdosing) the MTD. Despite both being considered unfavourable outcomes, the overdose is perceived as posing a greater risk to patients in terms of safety. Therefore, most of the model-based extensions implemented mechanisms to control the probability of overdosing, incorporating the concept of overdose control (EWOC), first introduced by Babb et al. in 1998[6] for the CRM model. However, even if the incorporation of the EWOC results in more conservative designs, it has been declared that the overdose control with the EWOC principle in several situations tends to be “excessively conservative”[20] often identifying subtherapeutic doses as MTD; on the other hand, in situations where the dose escalation is fast, the proportion of patients treated with toxic doses still remains higher adopting the model-based designs than the 3+3 design[21].

To address these issues, in 2022 Zhang et al.[22] proposed another escalation rule, applied to the BLRM, that accounted for both the underdosing and overdosing probability and improved the efficiency in selecting the correct MTD. Furthermore, in 2023 Ghosh et al.[24] proposed an alternative model-based approach called hybrid CRM (hCRM), allowing overdose control by implementing separate dose modification rules (e.g., for termination of the current and higher dose levels) based on an independent non-parametric approach at the current dose, thus reducing the assignment of overly toxic doses to the patients.

All the approaches above use only dose-limiting toxicities (DLTs) to guide dose escalation. According to the CTCAE, DLTs are grade 3 or higher adverse events. To improve dose-finding and avoid overly toxic doses, some methods have incorporated moderate toxicities (grades 1–2), referred to here as non-DLT adverse events (nDLTAEs). Some authors summarised different toxicity grades into a continuous response measure[29,30] or used multivariate ordinal models to account for them[31,32]. Others converted observed toxicities into a single metric, such as the ‘equivalent toxicity score’ or ‘toxicity burden’ [30,33]. These methods, however, often lack clinical interpretability due to the difficulty of linking combined outcomes to original toxicity grades. To address this, extensions of the Continual reassessment method (CRM) have been proposed. Lee et al.[26] in 2011 extended CRM to include both DLT and nDLTAE probabilities in defining the MTD, and later expanded it to cumulative toxicities[27] in 2019. Iasonos et al.[25] in 2011 also proposed a two-stage CRM Design, keeping DLTs as the main outcome, while incorporating moderate toxicities either as auxiliary outcomes in the first stage or within an ordinal outcome that combines both toxicity levels.

In this work, building on the state-of-the-art methods discussed above, we started from the BLRM, using the escalation rules proposed by Zhang et al.[22] and incorporated nDLTAE as an auxiliary outcome by adding an ad-hoc parameter to burden the DLT probability estimates, which remains the primary outcome for identifying the MTD.

In such an approach, clinicians are directly involved in defining (before the trial begins) and reporting (during the trial's conduct) the number of nDLTAEs that, although not meeting the definition of DLT, suggest that higher doses are very likely to result in DLTs. The more nDLTAEs are reported, the heavier the DLT probability estimates become. We defined this new version of BLRM as the Burdened bayesian logistic regression model (BBLRM).

The aim is to tackle two key issues that hinder the broader use of model-based approaches: clinicians' ongoing perception of limited control over dose-escalation decisions, and the challenge of minimizing the number of patients exposed to toxic doses without compromising the accuracy of identifying the true MTD due to excessive conservatism.

The paper is mainly focused on model-based designs and is organised as follows: in section 2.1, we recall the original BLRM with both overdose and underdose control proposed by Zhang et al.[22] (referred as Zhang's BLRM in the following). In section 2.2, we extend the BLRM to the BBLRM, defining in detail the ad-hoc nDLTAE parameter and describing precisely how it increases the estimates of DLT probabilities. In section 3, we describe the simulation setting adopted, focusing on the calibration phase involving two tuning parameters. In section 4, we provide a simulation study to compare the performance of BBLRM with three other model-based approaches: Zhang's BLRM[22], the hybrid CRM[24] and the two-stage CRM[25]. Finally, in section 5, the advantages and the critical aspects that need to be considered when applying this methodology are discussed.

## 2. Methods

### 2.1 Escalation Rules: BLRM with overdose and underdose control

A standard assumption underlying common cancer therapies (e.g. cytotoxicity therapies) is the monotonicity, stating that increasing toxicity increasing the anti-tumour activity of the compound. This assumption could not be valid for the most recent advanced therapies, where the toxicity-efficacy relationship could not be linear; however, in this work, we assume monotonicity. Considering the ethical constraints that must be respected during all the clinical trial phases, these constraints become even more stringent in oncology trials since, in every case, we observed adverse events as well as benefits. For these reasons, the aim of a phase I oncology trial is to minimise both the number of patients treated at low, non-therapeutic doses as well as the number given severely toxic overdoses. In such a safety concern mindset, Babb et al. in 1998[6] proposed the dose escalation with overdose control, a way to select a dose level for each cohort so that the predicted probability that the dose exceeds the MTD is less than or equal to a specified value $\gamma$. Let's consider two doses, $d_{min}, d_{max}$, assuming that the dose vector $D = \{d_1, d_2, \dots, d_i, \dots, d_I\}$ available for the study belongs to the interval $(d_{min}; d_{max})$. $y_i$ are realisations of $Y_i$, the DLT binary indicator variable related to the dose $d_i$ observed in a specific cohort $i$, such that $Y_i \sim Bin(n_i, p_i)$, where $n_i$ is the number of patients treated at the dose $d_i$ and $p_i$ is the DLT probability referred to the dose $d_i$. In this work, we considered the BLRM to model the dose-toxicity curve as follows:

$$log\left(\frac{p_i}{1-p_i}\right) = \theta_1 + e^{\theta_2} \cdot log\left(\frac{d_i}{d_{Ref}}\right)$$

where $d_{Ref}$ is any reference dose belonging to the dose vector. In this way, $\theta_1$ is the log-odds of the DLT probability associated with $d_{Ref}$; the exponential transformation $e^{\theta_2}$ allows $\theta_2 \in (-\infty; +\infty)$ and at the same time maintains the monotonicity assumption of the dose-toxicity curve. Since we are in a bayesian context, we can put a prior distribution (usually weakly informative) on the parameter $\theta = (\theta_1; \theta_2)$ and considering as $D_i$ the data collected up to the cohort $i$ we can obtain the posterior distribution for the parameters of interest $p(\theta_1, \theta_2|D_i)$. Considering $d_{MTD}$ the maximum tolerated dose, from the above posterior distribution, we can derive the marginal posterior distribution function of the MTD given $D_i$, $p(d_{MTD}|D_i)$, through the transformation formula and some algebra (see Babb et al.[6] for details). Therefore, the marginal posterior cumulative distribution is:

$$\pi_{D_i}(z) = \int_{d_{min}}^{z} p\,(d_{MTD}|D_i) d(d_{MTD}).$$

From this distribution, at each cohort, EWOC suggests as the next dose the one for which, based on all available data, the posterior probability of exceeding the MTD is equal to the feasibility bound $\gamma$. Therefore, the dose suggested for the next cohort $i+1$ is:

$$d_{i+1} = \pi_{D_i}^{-1}(\gamma)$$

However, since the dose suggested in this way can vary in a continuous range between $d_{min}$ and $d_{max}$ but, in practice, we have only a predefined discrete vector of doses, the real dose selected at the cohort $i$ is the closest to the one suggested originally. Therefore, the recommended dose at each cohort minimises the risk with respect to an asymmetric loss function:

$$l_\gamma(d_i, d_{MTD}) = \begin{cases} (1-\gamma) \cdot |d_k - d_{MTD}| & \text{if } d_i < d_{MTD} \\ \gamma \cdot |d_k - d_{MTD}| & \text{if } d_i > d_{MTD} \end{cases}$$

where $d_i$ is the dose selected for the cohort $i$ and $d_{MTD}$ is the true MTD. To apply this approach, we need to choose a value for the feasibility bound $\gamma$, that the authors suggest being in $(0; 0.5)$. With this range, we can easily see that the case of underdosing is heavily penalised with respect to the overdosing one, with equal distance $|d_i - d_{MTD}|$. This formulation makes EWOC overly conservative, especially when applied to BLRM, defining it as excessively cautious compared to simpler rule-based designs like the 3+3 method[22].

To overcome this issue and increase the performance of BLRM, Zhang et al. in 2022[22] proposed some complementary escalation rules that compare both the overdosing and underdosing probabilities. The purpose is to escalate the dose for the next cohort if, referred to the current dose, a portion of the underdosing probability is sufficiently higher than a portion of the overdosing probability. This ’portion’ can be interpreted as the feasibility bound of the EWOC. In practice, this has been developed in two different escalation rules. In line with the results and indications provided in Zhang et al.[22], in this work we adopted the escalation rule reported below:

*Escalation rule:*

Considering $\alpha \in (0; 1)$ the feasibility bound, the current dose $d_i$ is escalated to the subsequent dose $d_{i+1}$ in the next cohort if:

$$(1-\alpha) \cdot P_i(Under) > \alpha \cdot P_i(Over),$$

where $P_i(Under)$ and $P_i(Over)$ are the underdosing and overdosing probabilities for the dose $d_i$ respectively. If we look at the two probabilities mentioned above, these can vary in relation to the intervals that define a dose as an overdose, underdose and target dose. Such intervals are related to the toxicity probability $p_i$ for a generic dose $d_i$ defined as follows:

$$\begin{cases} \text{if } p_i \in (0; u) & \text{then } p_i \text{ is an underdose} \\ \text{if } p_i \in (u; o) & \text{then } p_i \text{ is the target dose} \\ \text{if } p_i \in (o; 1) & \text{then } p_i \text{ is an overdose} \end{cases}$$

Therefore, $P_i(Under) = P\big(p_i \in (0; u)\big)$ and $P_i(Over) = P\big(p_i \in (o; 1)\big)$.

The escalation rule is applied following these steps at each cohort for the dose suggestion:

1. Update the DLT probability estimates with all available data, thus underdosing and overdosing probabilities for each dose.
2. Apply first the escalation rule 1 of Zhang et al.[22]
3. If the condition of the escalation rule 1 is not met, then apply the EWOC principle.

An important note is related to the feasibility bound $\alpha$. Its value needs to be preliminarily chosen in order to check the conservative level of the escalation rule: the higher it is, the higher is the probability to escalate the dose in the next cohort, since the underdosing probability is more weighted with respect to the overdose one. Generally, the authors suggest that setting $\alpha$ to less than 0.5 can strike a good balance by enhancing the escalation's performance without making it too aggressive. In practice, the specific value of $\alpha$ needs to be chosen via a preliminary calibration phase. This aspect will be discussed in the simulations section.

## 2.2 BBLRM: Burdening the DLT probability using non-DLT Adverse Events

As mentioned in the introduction, despite the development of several model-based approaches in recent decades, clinicians remain hesitant to adopt these designs. In response to this, we proposed an extension of the BLRM with a primary focus on enhancing the model for estimating the DLT probability, rather than modifying the escalation rules. This approach aims to achieve two main objectives: first, to increase clinician involvement in the dose-escalation process within a model-based framework; and second, to improve the performance of the BLRM by reducing the allocation of toxic doses mistakenly identified as MTD, without sacrificing accuracy in selecting the true MTD. For escalation rules, we adopted the framework provided by Zhang et al.[22] and applied the specific steps outlined at the end of the previous section.

We introduced, then, an additional parameter $\delta$ to incorporate and account for the 'non-DLT adverse events' (nDLTAE) representing moderate toxicities and that, despite not meeting the study definition of DLT according to the criteria defined in the specific trial, suggests that higher doses are very likely to result in DLTs. For this reason, they can give more information about the dose-toxicity relationship. Moreover, since nDLTAEs are commonly reported along the study duration, their inclusion in the model does not increase the effort in data collection during the trial; on the other hand, they must be appropriately defined in the protocol before the beginning of the trial.

In Phase I oncology trials, rules for identifying DLTs are typically established based on the severity of adverse events and the mechanism of action of the investigational drug. These definitions often rely on the CTCAE grading system, which classifies AEs on a scale from Grade 1 (mild) to Grade 5 (death related to AE). However, the determination of whether a specific AE constitutes a DLT is not automatically derived from its CTCAE grade; instead, DLTs are defined a priori in the study protocol. For instance, Grade 3 or 4 toxicities such as febrile neutropenia, prolonged thrombocytopenia, or severe gastrointestinal events are commonly considered DLTs if they occur within a predefined window, typically the first treatment cycle (e.g., Days 1–28), when acute toxicities are most likely to emerge.

Our proposal builds on this framework by advocating for the prospective definition of a second category of adverse events: nDLTLAEs. These are adverse events that, while not meeting the criteria for DLTs, suggest a high likelihood of DLT occurrence at higher doses. For example, if a DLT is defined as a Grade 3 AE persisting for more than three days, then a Grade 3 AE lasting two days could be pre-specified as an nDLTLAE. Similarly, a persistent Grade 2 transaminitis might not qualify as a DLT but could serve as an early warning of potential Grade 3–4 hepatotoxicity at higher doses. A comparable case would be Grade 2 stomatitis that might progress to Grade 3 mucositis with dose escalation.

By involving clinicians more directly in the early stages of trial design, particularly in the identification and definition, a priori, of these predictive but non–dose-limiting toxicities, we aim to enhance the clinical relevance of the model-based dose-escalation process. This approach enables statistical models to incorporate nuanced toxicity signals that might otherwise be overlooked, thereby improving dose selection and patient safety.

Coming to the model specification, in the BLRM formula, $log\left(p_i/(1-p_i)\right) = \theta_1 + e^{\theta_2} \cdot log\left(d_i/d_{Ref}\right)$, the $\delta$ parameter is used to increment the posterior DLT probabilities estimated by the model, based on the cumulative count of patients for which any nDLTAEs are observed. Practically, after obtaining posterior estimates of the parameters $(\theta_1;\ \theta_2)$ via MCMC sampling, $\delta$ is added to proportionally adjust the value of $\theta_1$, thereby increasing the DLT toxicity estimates produced by the original BLRM. Through different ways to incorporate $\delta$, we found that burdening $\theta_1$, instead of $\theta_2$, provided more stable parameter estimates, resulting in a consistent burdening effect on DLT probability estimates across various doses. Indeed, since $\theta_2$ is directly linked to the dose, increasing it proportionally with the number of patients who experienced any nDLTAEs resulted in a significantly variable impact on the DLT probabilities across the dose vector. Conversely, increasing $\theta_1$ resulted in a more uniform and consistent burdening effect of DLT probabilities across the different doses.

Formally, the Burdened bayesian logistic regression model is expressed as:

$$log\left(\frac{p_i}{1-p_i}\right) = \theta_1 + |\delta \cdot \theta_1| + e^{\theta_2} \cdot log\left(\frac{d_i}{d_{Ref}}\right),$$

where the absolute value ensures that DLT probabilities increase as $\delta$ grows.

At each cohort, after having collected all the available information related to DLT and nDLTAEs, $\delta$ is generated from a uniform random variable defined as follows, with a range determined by the current count of patients for which at least one nDLTAE was observed:

$$\delta \sim U\left(max\left(0, \omega\frac{s_{i^*}-1}{n_{i^*}}\right); \omega\frac{s_{i^*}}{n_{i^*}}\right).$$

The upper bound of the Uniform is linked to the number of patients for whom any nDLTAES were observed up to the cohort $i$ $(s_{i^*})$ related to the overall number of patients treated up to the cohort $i$ $(n_{i^*})$, while the lower bound is set to $\frac{s_{i^*}-1}{n_{i^*}}$, to limit the range in which $\delta$ can be generated but maintaining at the same time a source of variability, constraining it not to be lower than zero in case of none experiences a nDLTAE.

Within the Bayesian framework, it is important to note that $\delta$ does not serve the same role as the primary model parameters $(\theta_1, \theta_2)$ as it is neither inferred jointly with them nor included in the formal posterior distribution. Rather, $\delta$ is introduced as an externally generated burdening factor to reflect the number of patients experiencing nDLTAEs. Its purpose is to increase the posterior estimates of $\theta_1$, thereby increasing the estimated DLT probabilities. Consequently, the model departs from a strictly Bayesian formulation by incorporating elements of empirical adjustment. In modelling nDLTAEs, it is essential to incorporate cumulative patient data across all cohorts, rather than relying solely on observations from the current cohort. Restricting the $\delta$ generation considering only the current cohort introduces a memoryless process into the model; for instance, if no nDLTAEs are observed in the present cohort despite prior occurrences, the parameter $\delta$ would collapse to $\sim U(0; 0)$, effectively discarding previously accrued evidence. By defining $\delta$ as a function of the cumulative proportion of patients experiencing nDLTAEs relative to the total number treated, the model achieves a more accurate and robust representation of the evolving toxicity profile. Importantly, from a clinical perspective, this approach aligns with how safety signals are interpreted in practice: the risk associated with treatment should reflect not only immediate cohort outcomes but also the broader trial experience, as the original BLRM does with the DLT observed. The incorporation of $\delta$ into the estimation of dose-limiting toxicity (DLT) probabilities thus provides a dynamic and interpretable framework, wherein the influence of previous nDLTAEs on DLT probability estimates evolves proportionally with cumulative trial data.

Focusing on the expression that defines the generation of $\delta$, we observed that it can vary between 0 and 1, with specific sub-intervals determined by the proportion of patients who have experienced at least one nDLTAE up to the current cohort. Exploratory simulations revealed that allowing $\delta$ to span this full range resulted in an excessive burdening effect on $\theta_1$, leading to inflated toxicity estimates and overly conservative dose escalation decisions.

To address this, we introduced a tuning parameter $\omega \in (0,1)$, which narrows the boundaries of the Uniform distribution from which $\delta$ is drawn. This parameter should be calibrated prior to the start of the trial, based on trial-specific factors such as cohort size and the available dose levels, as well as model-specific components like the prior distributions for $(\theta_1; \theta_2)$, and the choice of reference dose $d_{Ref}$. Notably, higher values of $\omega$ result in stronger burdening effects from $\delta$, yielding more conservative dose escalation behaviour. The calibration of $\omega$ should be conducted in conjunction with the escalation threshold parameter $\alpha$, as discussed in the following section. Furthermore, the burdening effect on the DLT probability of $\delta$ and $\omega$ parameters is shown at the end of the two tuning parameters $(\alpha; \omega)$ calibration's process in section 3.2.

## 3. Simulations

### 3.1 Simulation settings

The simulation study conducted aims to evaluate the performance of the BBLRM in comparison with different alternatives: Zhang's BLRM[22], used as our reference model, and two other approaches introduced earlier: the two-stage Continual reassessment method (CRM) (Iasonos et al.[25]) and the hybrid Continual reassessment method (hCRM) proposed by Ghosh et al.[24]; these models align with our goals of incorporating nDLTAEs and/or minimizing the allocation of toxic doses selected as MTD.

As a motivating example for the simulation study, we referred to the phase I first-in-human trial described in Sessa et al.[23] in 2013. The original trial utilised the Bayesian logistic regression model (BLRM) for dose escalation. Nine doses were considered, represented by the vector $D = \{2,4,8,16,22,28,40,54,70\}$. Up to 27 patients were treated in cohorts of three, with each cohort assigned to the same dose. For the prior distribution of the BLRM parameters $(\theta_1;\ \theta_2)$, we adopted a weakly informative bivariate normal distribution with a mean of $(-1.099;\ 0)$, standard deviations of $(2;\ 1)$, and a correlation coefficient of 0. These specific values are taken from the work of Zheng and Hampson[15], which considered the same trial of Sessa et al.[23] as a motivating example.

For the reference dose $d_{Ref}$, we primarily selected two dose levels: the middle dose (22) and the highest dose (70). In both cases, $d_{Ref}$ is interpreted as the expected MTD according also with the prior distribution set for $(\theta_1;\ \theta_2)$ as described by Zheng and Hampson[15]. For completeness and to further assess model performance, we also considered setting $d_{Ref}$ at the lowest dose (2) in the dose vector D; however, this choice is less realistic, especially in the context of a nine-dose sequence. Therefore, for the sake of clarity and brevity, we report the results for $d_{Ref} = \{22{,}70\}$ in the main manuscript and include those for $d_{Ref} = 2$ in the supplementary material.

These settings were applied to the BBLRM, the BLRM with Zhang's escalation rule, and the hCRM, which also relies on the BLRM for DLT probability estimation.

To apply the escalation rule of Zhang et al.[22] adopted in Zhang's BLRM and BBLRM, we set $(0;\ 0.16)$ and $(0.33;\ 1.00)$ as the underdosing and overdosing probability intervals related to the DLT probability as defined in section 2.1; the target DLT probability for the MTD is set to 0.25.

An essential aspect of the simulation study was defining how to generate DLT and nDLTAE events.

In the context of common standard therapies (e.g., cytotoxic therapies), the relationship between DLTs and nDLTAEs tends to be relatively predictable and dose-dependent, highlighting how important it is to consider the occurrence of early lower signals of toxicities, such as nDLTAEs, to inform and possibly prevent the occurrence of DLT in subsequent doses. Examples of such events are myelosuppression, mucositis, and gastrointestinal symptoms. These toxicities typically increase in frequency and severity with higher doses, making lower-grade events or nDLTAEs (e.g., Grade 1–2 neutropenia or mucositis) useful predictors of subsequent DLTs or considering cumulating safety (Seymour & Eisenhauer[35] 2001; Postel-Vinay et al.[36] 2014).

Given these considerations, we formulated two main relationships between DLT and nDLTAEs probabilities defined below, which can mainly reflect the clinical practice:

Relation 1:

In this relation, we assumed that when a dose has a DLT probability over the target ($P(DLT) > 0.25$); then, it's very plausible to observe at least an nDLTAE. Indeed, P(nDLTAE) faster increases as P(DLT) increases up to the DLT target toxicity is reached, then a plateau equal to 0.95 is set for P(nDLTAE) when $P(DLT) = 0.33$, the lower bound of the overdosing probability interval.

$$P(nDLTAE) = \begin{cases} 0.10 & \text{if } P(DLT) \in (0; 0.05) \\ 0.20 & \text{if } P(DLT) \in [0.05; 0.10) \\ 0.35 & \text{if } P(DLT) \in [0.10; 0.15) \\ 0.55 & \text{if } P(DLT) \in [0.15; 0.20) \\ 0.85 & \text{if } P(DLT) \in [0.20; 0.25) \\ 0.90 & \text{if } P(DLT) \in [0.25; 0.30) \\ 0.93 & \text{if } P(DLT) \in [0.30; 0.33) \\ 0.95 & \text{if } P(DLT) \in [0.33; 1.00) \end{cases}$$

Relation 2:

In this case, we assume that when a dose has a DLT probability over the target ($P(DLT) > 0.25$), P(nDLTAE) progressively decreases, reflecting the scenario where all potential adverse events are severe enough to be classified as DLTs. Thus, P(nDLTAE) increases in the same way as in relation 1, up to the DLT target toxicity is reached, then progressively decrease up to a low constant probability of 0.10.

$$P(nDLTAE) = \begin{cases} 0.10 & \text{if } P(DLT) \in (0; 0.05) \\ 0.20 & \text{if } P(DLT) \in [0.05; 0.10) \\ 0.35 & \text{if } P(DLT) \in [0.10; 0.15) \\ 0.55 & \text{if } P(DLT) \in [0.15; 0.20) \\ 0.85 & \text{if } P(DLT) \in [0.20; 0.25) \\ 0.65 & \text{if } P(DLT) \in [0.25; 0.30) \\ 0.50 & \text{if } P(DLT) \in [0.30; 0.33) \\ 0.25 & \text{if } P(DLT) \in [0.33; 0.40) \\ 0.10 & \text{if } P(DLT) \in [0.40; 1.00) \end{cases}$$

The two relations are summarised in Figure 1.

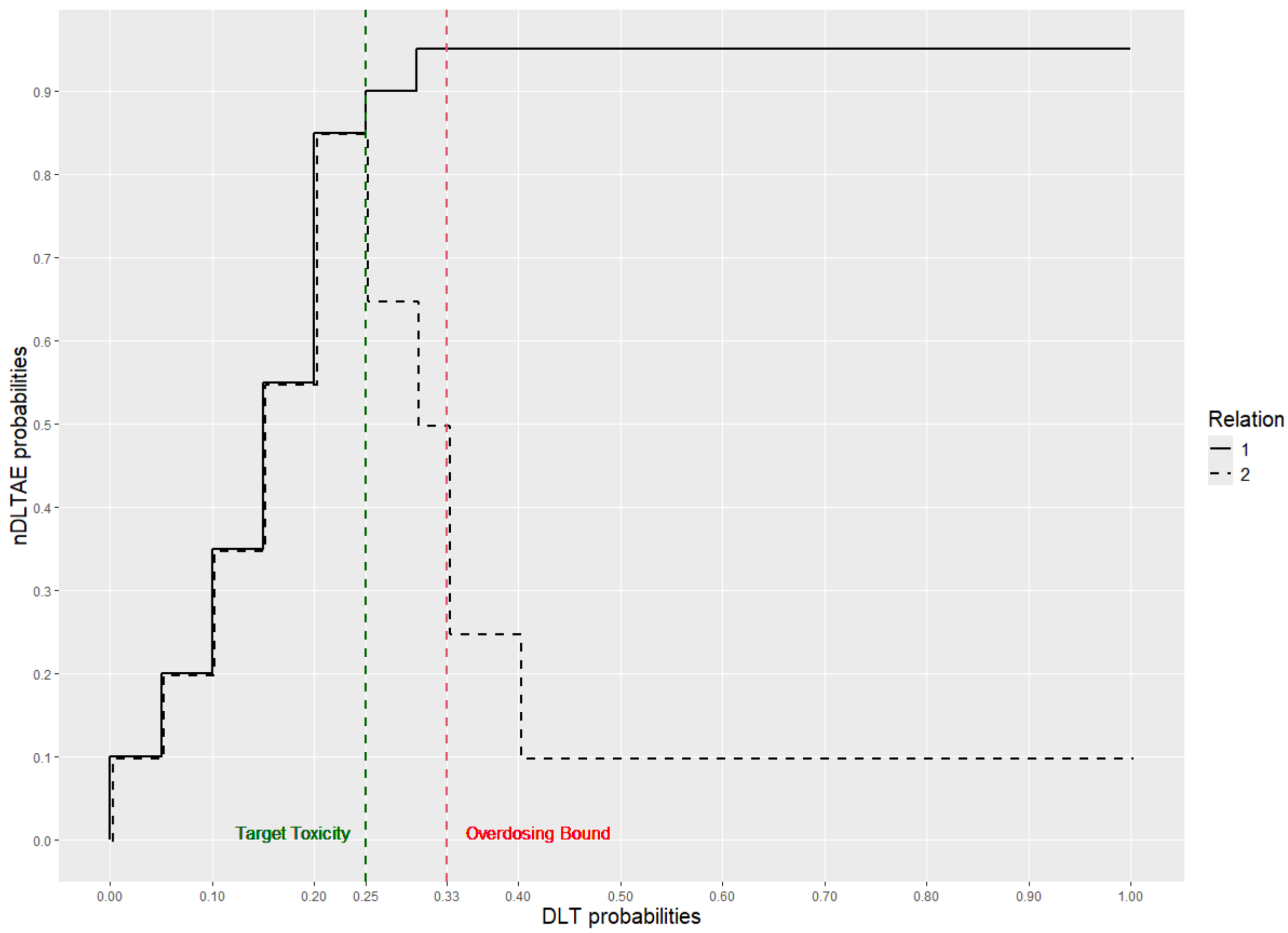

*Figure 1: Probability of nDLTAE for a generic dose based on the probability of DLT.*

Moreover, to stress the model and check the performance, we considered two further relationships between DLT and nDLTAEs: the first one is a more extreme case of Relation 2, where P(nDLTAE) decreasing even faster when the DLT probability is above the target rate; the second one addresses the atypical situations where the nDLTAE probability is constant at each dose, regardless of P(DLT). Detailed descriptions and models' results of these two further cases are reported in the supplementary material.

Regarding the DLT probability, for the same dose vector $D = \{2,4,8,16,22,28,40,54,70\}$, nine different scenarios (i.e., $S_1, S_2, \ldots, S_9$) have been considered, varying the MTD ($P(DLT = 0.25)$) across all the doses belonging to the dose vector.

In Table 1, the DLT probability for each scenario is defined. From that, the respective nDLTAE probability vector is derived for Relation 1 and Relation 2 and reported in Table 2 and Table 3 respectively.

*Table 1: DLT probabilities for the nine scenarios.*

| Dose vector and DLT probability | | | | | | | | | |
|---|---|---|---|---|---|---|---|---|---|
| Dose vector | 2 | 4 | 8 | 16 | 22 | 28 | 40 | 54 | 70 |
| $Sc_1$ | 0.25 | 0.35 | 0.41 | 0.47 | 0.52 | 0.58 | 0.63 | 0.70 | 0.78 |
| $Sc_2$ | 0.11 | 0.25 | 0.35 | 0.41 | 0.47 | 0.52 | 0.58 | 0.63 | 0.70 |
| $Sc_3$ | 0.08 | 0.16 | 0.25 | 0.35 | 0.42 | 0.45 | 0.53 | 0.60 | 0.70 |
| $Sc_4$ | 0.02 | 0.05 | 0.14 | 0.25 | 0.35 | 0.42 | 0.51 | 0.60 | 0.68 |
| $Sc_5$ | 0.03 | 0.05 | 0.10 | 0.16 | 0.25 | 0.35 | 0.40 | 0.48 | 0.55 |
| $Sc_6$ | 0.001 | 0.005 | 0.03 | 0.10 | 0.16 | 0.25 | 0.38 | 0.50 | 0.60 |
| $Sc_7$ | 0.01 | 0.02 | 0.05 | 0.08 | 0.11 | 0.14 | 0.25 | 0.37 | 0.47 |
| $Sc_8$ | 0.01 | 0.03 | 0.04 | 0.05 | 0.08 | 0.11 | 0.14 | 0.25 | 0.37 |
| $Sc_9$ | 0.01 | 0.03 | 0.04 | 0.05 | 0.08 | 0.11 | 0.14 | 0.18 | 0.25 |

*Table 2: nDLTAE probabilities for the nine scenarios (relation 1).*

| **Dose vector and nDLTAE probability** | | | | | | | | | |
|---|---|---|---|---|---|---|---|---|---|
| Dose vector | 2 | 4 | 8 | 16 | 22 | 28 | 40 | 54 | 70 |
| $Sc_1$ | 0.85 | 0.95 | 0.95 | 0.95 | 0.95 | 0.95 | 0.95 | 0.95 | 0.95 |
| $Sc_2$ | 0.35 | 0.85 | 0.95 | 0.95 | 0.95 | 0.95 | 0.95 | 0.95 | 0.95 |
| $Sc_3$ | 0.20 | 0.55 | 0.85 | 0.95 | 0.95 | 0.95 | 0.95 | 0.95 | 0.95 |
| $Sc_4$ | 0.10 | 0.20 | 0.35 | 0.85 | 0.95 | 0.95 | 0.95 | 0.95 | 0.95 |
| $Sc_5$ | 0.10 | 0.20 | 0.35 | 0.55 | 0.85 | 0.95 | 0.95 | 0.95 | 0.95 |
| $Sc_6$ | 0.10 | 0.10 | 0.10 | 0.35 | 0.55 | 0.85 | 0.95 | 0.95 | 0.95 |
| $Sc_7$ | 0.10 | 0.10 | 0.20 | 0.20 | 0.35 | 0.55 | 0.85 | 0.95 | 0.95 |
| $Sc_8$ | 0.10 | 0.10 | 0.10 | 0.20 | 0.20 | 0.35 | 0.55 | 0.85 | 0.95 |
| $Sc_9$ | 0.10 | 0.10 | 0.10 | 0.10 | 0.20 | 0.20 | 0.35 | 0.55 | 0.85 |

*Table 3: nDLTAE probabilities for the nine scenarios (relation 2).*

| **Dose vector and nDLTAE probability** | | | | | | | | | |
|---|---|---|---|---|---|---|---|---|---|
| Dose vector | 2 | 4 | 8 | 16 | 22 | 28 | 40 | 54 | 70 |
| $Sc_1$ | 0.85 | 0.25 | 0.10 | 0.10 | 0.10 | 0.10 | 0.10 | 0.10 | 0.10 |
| $Sc_2$ | 0.35 | 0.85 | 0.25 | 0.10 | 0.10 | 0.10 | 0.10 | 0.10 | 0.10 |
| $Sc_3$ | 0.20 | 0.55 | 0.85 | 0.25 | 0.10 | 0.10 | 0.10 | 0.10 | 0.10 |
| $Sc_4$ | 0.10 | 0.20 | 0.35 | 0.85 | 0.25 | 0.10 | 0.10 | 0.10 | 0.10 |
| $Sc_5$ | 0.10 | 0.20 | 0.35 | 0.55 | 0.85 | 0.25 | 0.10 | 0.10 | 0.10 |
| $Sc_6$ | 0.10 | 0.10 | 0.10 | 0.35 | 0.55 | 0.85 | 0.25 | 0.10 | 0.10 |
| $Sc_7$ | 0.10 | 0.10 | 0.20 | 0.20 | 0.35 | 0.55 | 0.85 | 0.25 | 0.10 |
| $Sc_8$ | 0.10 | 0.10 | 0.10 | 0.20 | 0.20 | 0.35 | 0.55 | 0.85 | 0.25 |
| $Sc_9$ | 0.10 | 0.10 | 0.10 | 0.10 | 0.20 | 0.20 | 0.35 | 0.55 | 0.85 |

### 3.2 Tuning parameters' phase

In the BBLRM, two tuning parameters are employed: $\alpha$, associated with Zhang's escalation rule[22], and $\omega$, which influences the burdening effect of the parameter $\delta$. These parameters serve distinct but complementary functions in the model.

$\alpha$ reflects the weight given to underdosing and overdosing probabilities. Specifically, higher values of $\alpha$ increase the underdosing's weight, thereby encouraging dose escalation. Although $\alpha$ ranges within (0, 1), Zhang et al.[22] recommend not exceeding a value of 0.5 to avoid overly aggressive dose escalation.

Conversely, $\omega$ governs the boundaries of the Uniform distribution from which $\delta$ is generated for each cohort. A higher $\omega$ value expands these boundaries, amplifying the burdening effect in estimating the DLT probabilities. In our simulation study, $\omega$ varied between 0.30 and 0.60, combined with different values of $\alpha$.

The calibration process relies on two key operating characteristics: the proportion of toxic doses ($\mathrm{P(DLT)} > 0.33$) that are incorrectly selected as the MTD, and the proportion of true MTDs selected ($\mathrm{P(DLT)} = 0.25$).

The calibration process was conducted in two phases. First, we calibrated the α parameter alone to determine the optimal value for Zhang's BLRM, which served as our benchmark. Subsequently, for each value of $\alpha$ different values of $\omega$ were tested to calibrate the BBLRM, identifying the optimal pair of values for $(\alpha; \omega)$. All the models have been tested simulating 10000 trials for each scenario (i.e., $Sc_1, Sc_2, \dots, Sc_9$) and for each relation between $\mathrm{P(DLT)}$ and $\mathrm{P(nDLTAE)}$ described in section 3.1.

For brevity and clarity, the main manuscript presents the calibration process only for Relation 1, while the supplementary material includes the corresponding figures and tables for Relation 2. The different combinations of $(\alpha; \omega)$ considered in the calibration process for the dose reference $d_{ref} = \{22,70\}$ are reported in Table 4 below.

*Table 4: Combinations of (α, ω) values tested in the calibration phase.*

| | | **α** | | | |
|---|---|---|---|---|---|
| | | 0.25 | 0.30 | 0.35 | 0.40 |
| **ω** | 0.30 | x | x | | |
| | 0.35 | x | x | | |
| | 0.40 | x | x | | |
| | 0.50 | | | x | x |
| | 0.55 | | | x | x |
| | 0.60 | | | x | x |

The calibration process for $d_{ref} = 2$ is reported in the supplementary material as stated in section 3.1. As shown, some of the listed combinations were not tested. This is because a lower value of $\alpha$ combined with a higher value of $\omega$ led to overly conservative dose escalation, which reduced the accuracy in identifying the true MTD. Conversely, a higher $\alpha$ value combined with a lower $\omega$ value resulted in excessively aggressive dose escalation, increasing the likelihood of selecting toxic doses as the MTD.

In calibrating α for Zhang's BLRM, the model demonstrated comparable operating characteristics across the tested values, as shown in Figure 2 below. However, we selected $\alpha = 0.25$ as it yielded slightly better overall results across the various scenarios.

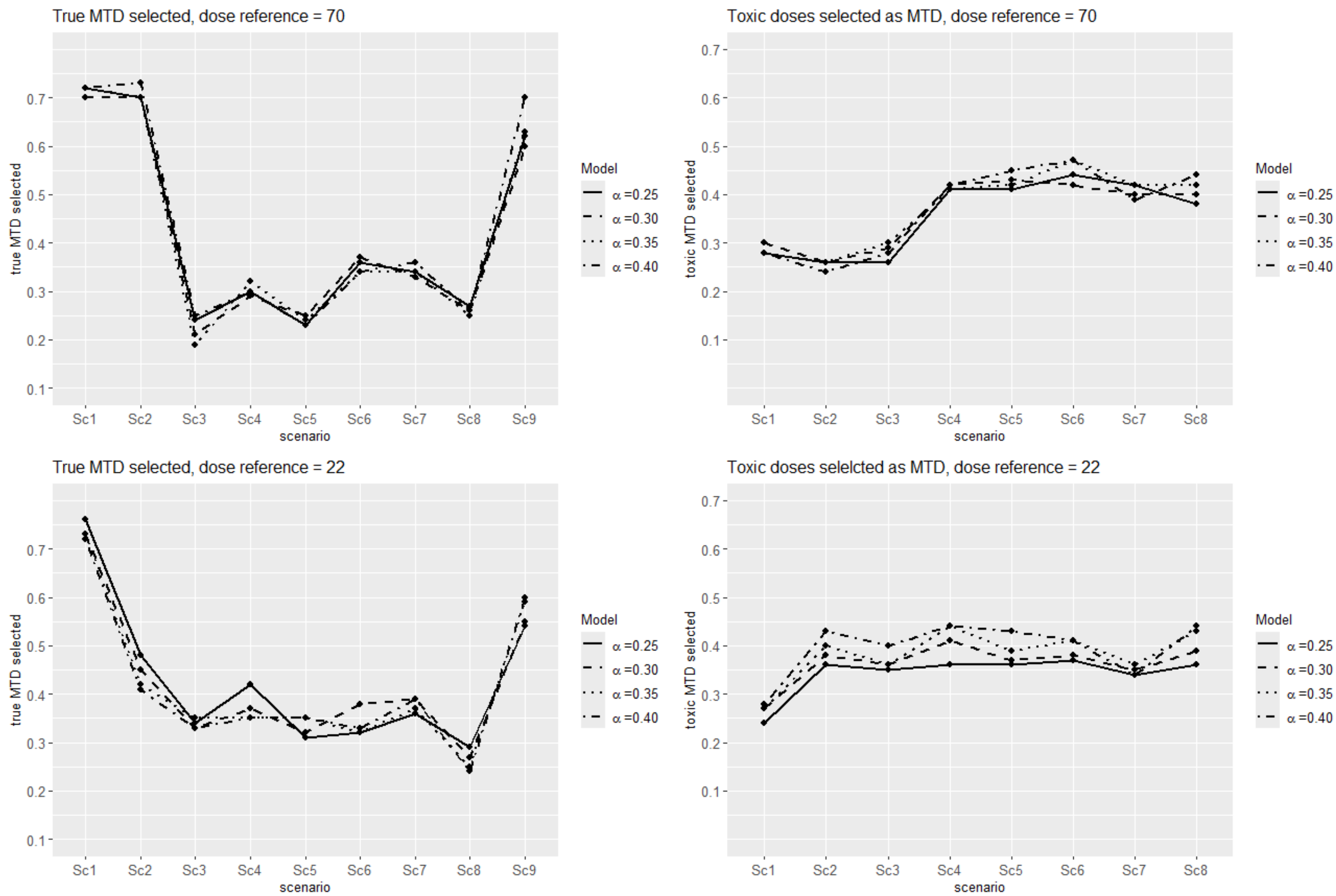

*Figure 2: Operating characteristics of the Zhang's BLRM for different values of the $\alpha$ parameter, with , $d_{ref} = 70$ and , $d_{ref} = 22$.*

Note: Since Zhang's BLRM does not account for nDLTAEs, its calibration focused only on the nine scenarios for DLT probability (Table 1). Additionally, scenario Sc9 was omitted from all plots related to the proportion of toxic doses selected as MTD, as the MTD corresponds to the last dose in the dose vector D, with no doses defined as toxic (i.e., P(DLT) $>$ 0.33; see Table 1, line 9).

For the calibration of the BBLRM, each combination of (α, ω) listed in Table 4 was evaluated. Results for Relation 1 are presented in Table 5 and Figures 3 and 4 for $d_{ref} = 70$, and in Table 6 and Figures 5 and 6 for $d_{ref} = 22$. In all figures, the solid lines represent the baseline BLRM model without burdening (i.e. ω = 0, Zhang's BLRM).

*Table 5: Summary statistics related to proportions of toxic doses wrongly selected as MTD (left panel) and true MTD selected (right panel), relation 1,* $d_{ref} = 70$.

| **Relation 1: toxic doses wrongly selected as MTD** | | | | |
|---|---|---|---|---|
| $(\alpha; \omega)$ | **Mean** | **Median** | **Min** | **Max** |
| (0.25 – 0.00)* | 36 % | 40 % | 26 % | 44 % |
| (0,25 – 0.30) | 30 % | 32 % | 17 % | 37 % |
| (0,25 – 0.35) | 29 % | 31 % | 18 % | 39 % |
| (0,25 – 0.40) | 26 % | 29 % | 14 % | 37 % |
| (0.30 – 0.00)* | 37 % | 40 % | 26 % | 43 % |
| (0,30 – 0.30) | 31 % | 34 % | 19 % | 42 % |
| (0,30 – 0.35) | 29 % | 31 % | 17 % | 41 % |
| (0,30 – 0.40) | 32 % | 29 % | 18 % | 40 % |
| (0.35 – 0.00)* | 37 % | 41 % | 26 % | 47 % |
| (0,35 – 0.50) | 28 % | 30 % | 15 % | 39 % |
| (0,35 – 0.55) | 29 % | 31 % | 15 % | 37 % |
| (0,35 – 0.60) | 25 % | 26 % | 14 % | 36 % |
| (0.40 – 0.00)* | 37 % | 40 % | 24 % | 47 % |
| (0,40 – 0.50) | 29 % | 31 % | 17 % | 44 % |
| (0,40 – 0.55) | 27 % | 30 % | 14 % | 40 % |
| (0,40 – 0.60) | 26 % | 28 % | 16 % | 35 % |

| **Relation 1: true MTD selected** | | | | |
|---|---|---|---|---|
| $(\alpha; \omega)$ | **Mean** | **Median** | **Min** | **Max** |
| (0.25 – 0.00)* | 42 % | 34 % | 23 % | 72 % |
| (0,25 – 0.30) | 44 % | 38 % | 26 % | 70 % |
| (0,25 – 0.35) | 43 % | 37 % | 25 % | 70 % |
| (0,25 – 0.40) | 43 % | 37 % | 24 % | 71 % |
| (0.30 – 0.00)* | 41 % | 33 % | 21 % | 70 % |
| (0,30 – 0.30) | 43 % | 36 % | 28 % | 68 % |
| (0,30 – 0.35) | 44 % | 38 % | 27 % | 71 % |
| (0,30 – 0.40) | 44 % | 38 % | 23 % | 69 % |
| (0.35 – 0.00)* | 41 % | 34 % | 19 % | 72 % |
| (0,35 – 0.50) | 43 % | 36 % | 24 % | 71 % |
| (0,35 – 0.55) | 41 % | 33 % | 25 % | 67 % |
| (0,35 – 0.60) | 43 % | 34 % | 24 % | 73 % |
| (0.40 – 0.00)* | 43 % | 34 % | 23 % | 73 % |
| (0,40 – 0.50) | 42 % | 37 % | 22 % | 69 % |
| (0,40 – 0.55) | 43 % | 36 % | 25 % | 69 % |
| (0,40 – 0.60) | 44 % | 37 % | 27 % | 71 % |

*Zhang's BLRM[22], the reference model.

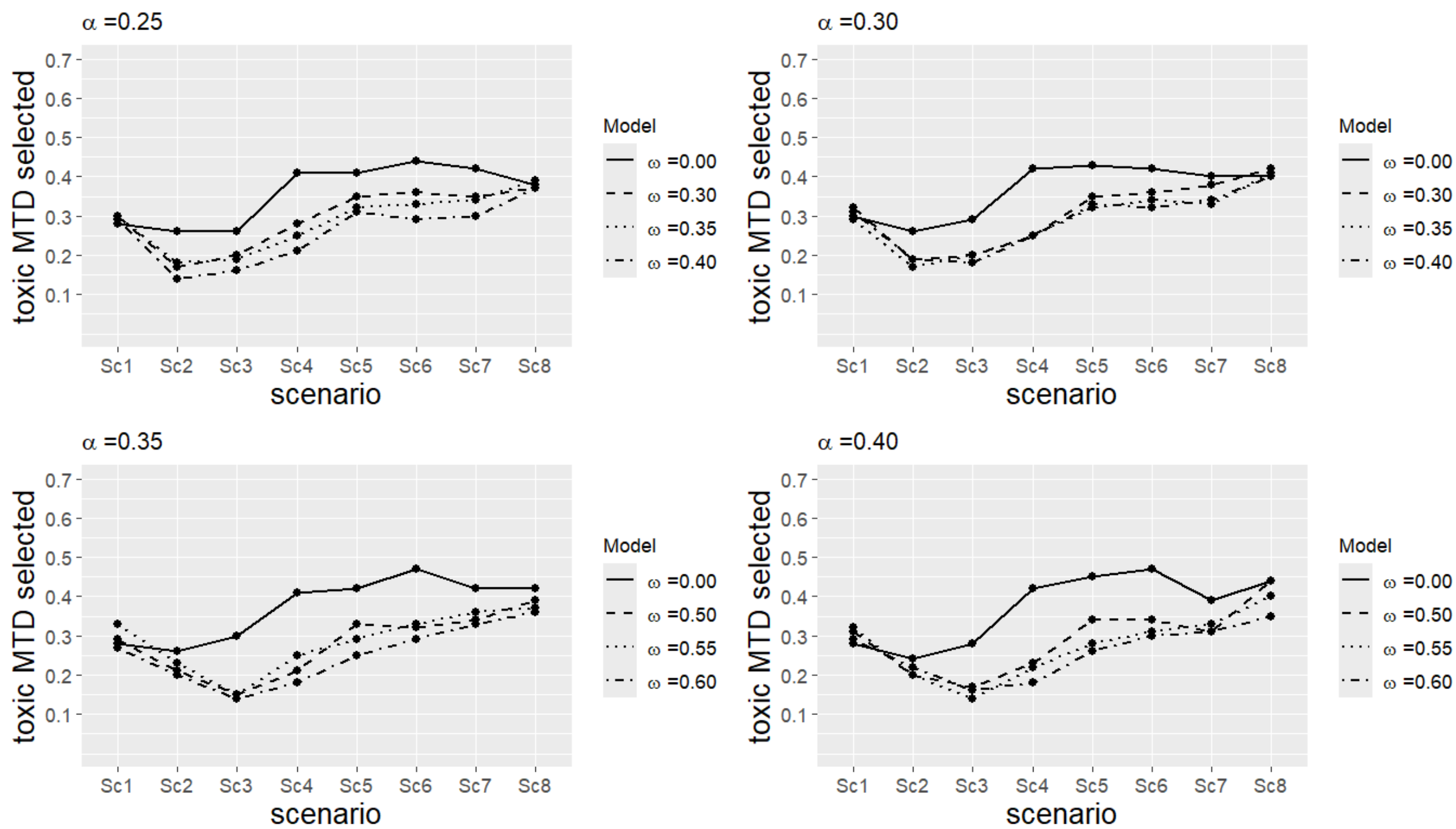

*Figure 3: Proportions of toxic doses selected as MTD for different combinations of (α, ω), relation 1,* $d_{ref} = 70$.

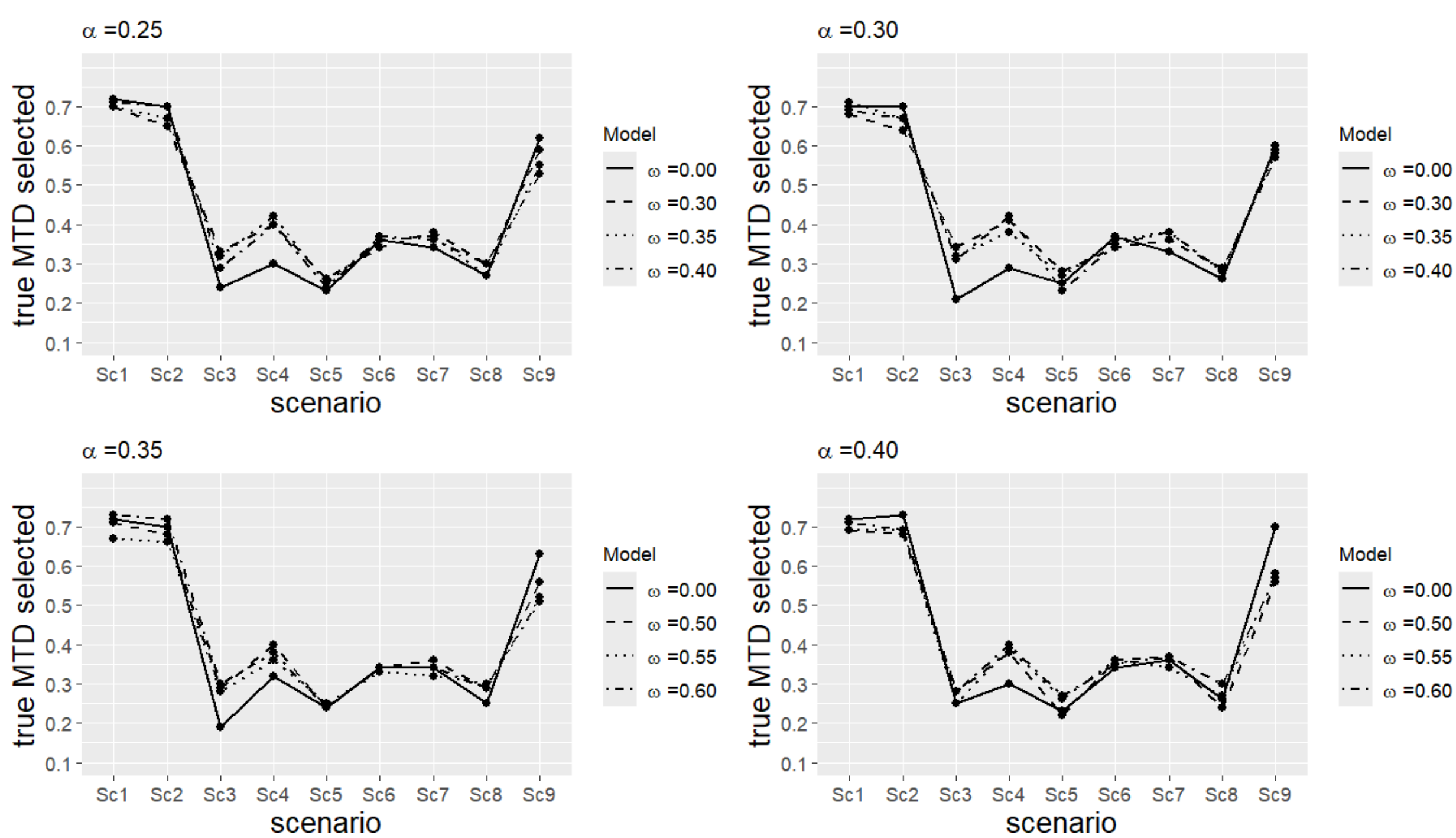

*Figure 4: Proportions of true MTD selected for different combinations of (α, ω), relation 1,* $d_{ref} = 70$.

*Table 6: Summary statistics related to proportions of toxic doses wrongly selected as MTD (left panel) and true MTD selected (right panel), relation 1,* $d_{ref} = 22$.

| Relation 1: toxic doses wrongly selected as MTD | | | | |
|---|---|---|---|---|
| **$(\alpha; \omega)$** | **Mean** | **Median** | **Min** | **Max** |
| (0.25 – 0.00)* | 0.34 | 0.36 | 0.24 | 0.37 |
| (0,25 – 0.30) | 0.24 | 0.25 | 0.18 | 0.28 |
| (0,25 – 0.35) | 0.23 | 0.23 | 0.17 | 0.30 |
| (0,25 – 0.40) | 0.20 | 0.20 | 0.15 | 0.19 |
| (0.30 – 0.00)* | 0.36 | 0.37 | 0.27 | 0.41 |
| (0,30 – 0.30) | 0.25 | 0.25 | 0.19 | 0.31 |
| (0,30 – 0.35) | 0.23 | 0.23 | 0.17 | 0.30 |
| (0,30 – 0.40) | 0.21 | 0.21 | 0.15 | 0.28 |
| (0.35 – 0.00)* | 0.38 | 0.40 | 0.27 | 0.44 |
| (0,35 – 0.50) | 0.20 | 0.20 | 0.13 | 0.32 |
| (0,35 – 0.55) | 0.19 | 0.19 | 0.13 | 0.31 |
| (0,35 – 0.60) | 0,17 | 0.16 | 0.09 | 0.29 |
| (0.40 – 0.00)* | 0.40 | 0.42 | 0.28 | 0.44 |
| (0,40 – 0.50) | 0.22 | 0.22 | 0.15 | 0.29 |
| (0,40 – 0.55) | 0.20 | 0.21 | 0.09 | 0.31 |
| (0,40 – 0.60) | 0.18 | 0.19 | 0.10 | 0.31 |

| Relation 1: true MTD selected | | | | |
|---|---|---|---|---|
| **$(\alpha; \omega)$** | **Mean** | **Median** | **Min** | **Max** |
| (0.25 – 0.00)* | 0.42 | 0.36 | 0.29 | 0.76 |
| (0,25 – 0.30) | 0.42 | 0.37 | 0.28 | 0.72 |
| (0,25 – 0.35) | 0.42 | 0.35 | 0.29 | 0.70 |
| (0,25 – 0.40) | 0.42 | 0.35 | 0.28 | 0.71 |
| (0.30 – 0.00)* | 0.42 | 0.38 | 0.27 | 0.73 |
| (0,30 – 0.30) | 0.43 | 0.41 | 0.28 | 0.69 |
| (0,30 – 0.35) | 0.44 | 0.41 | 0.29 | 0.70 |
| (0,30 – 0.40) | 0.44 | 0.37 | 0.27 | 0.72 |
| (0.35 – 0.00)* | 0.42 | 0.35 | 0.25 | 0.73 |
| (0,35 – 0.50) | 0.43 | 0.38 | 0.23 | 0.68 |
| (0,35 – 0.55) | 0.43 | 0.37 | 0.23 | 0.69 |
| (0,35 – 0.60) | 0.43 | 0.37 | 0.20 | 0.71 |
| (0.40 – 0.00)* | 0.41 | 0.35 | 0.24 | 0.72 |
| (0,40 – 0.50) | 0.42 | 0.37 | 0.21 | 0.71 |
| (0,40 – 0.55) | 0.44 | 0.41 | 0.20 | 0.69 |
| (0,40 – 0.60) | 0.43 | 0.37 | 0.18 | 0.69 |

*Zhang's BLRM[22], the reference model.

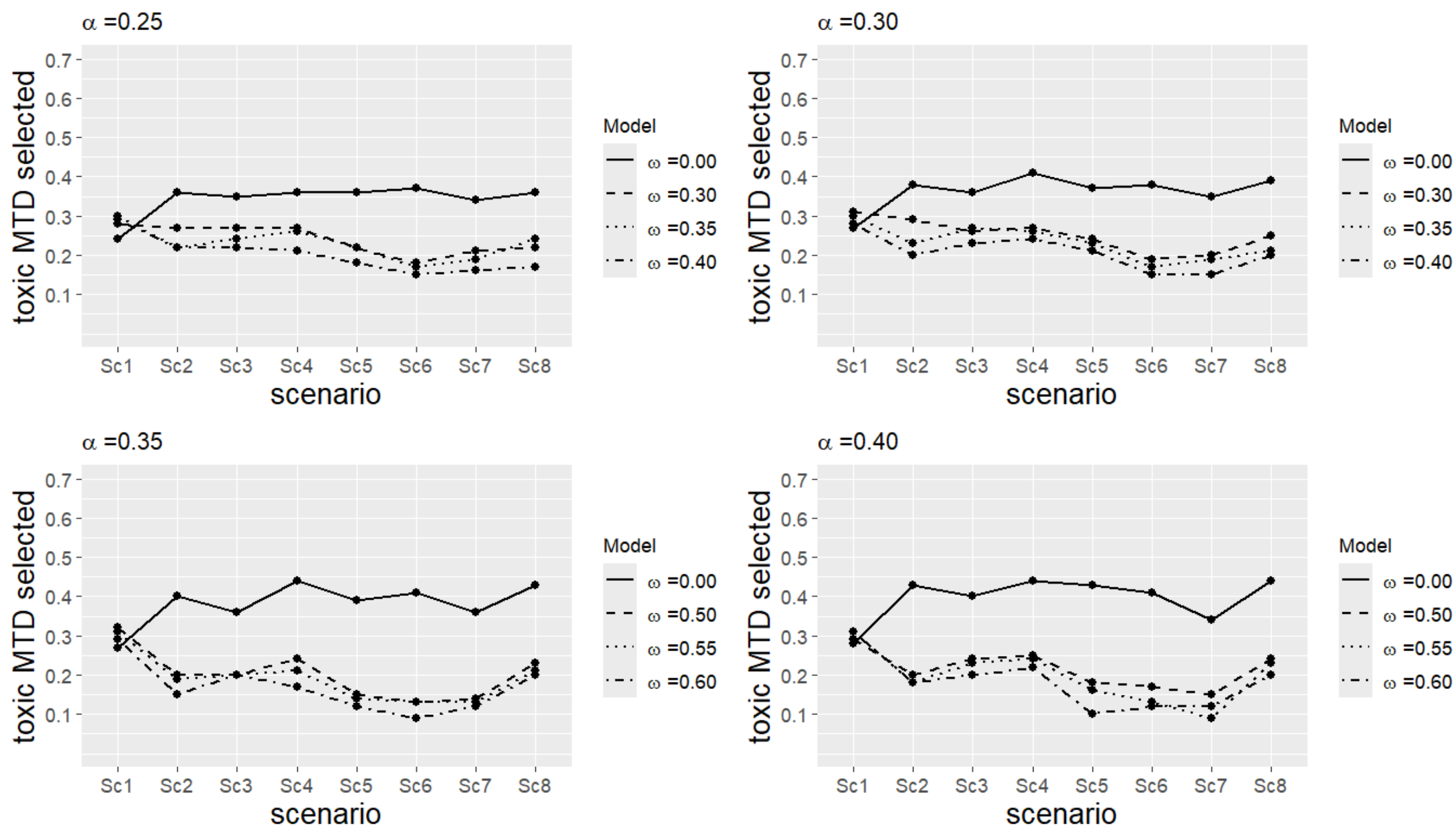

*Figure 5: Proportions of toxic MTD selected for different combinations of (α, ω), relation 1,* $d_{ref} = 22$.

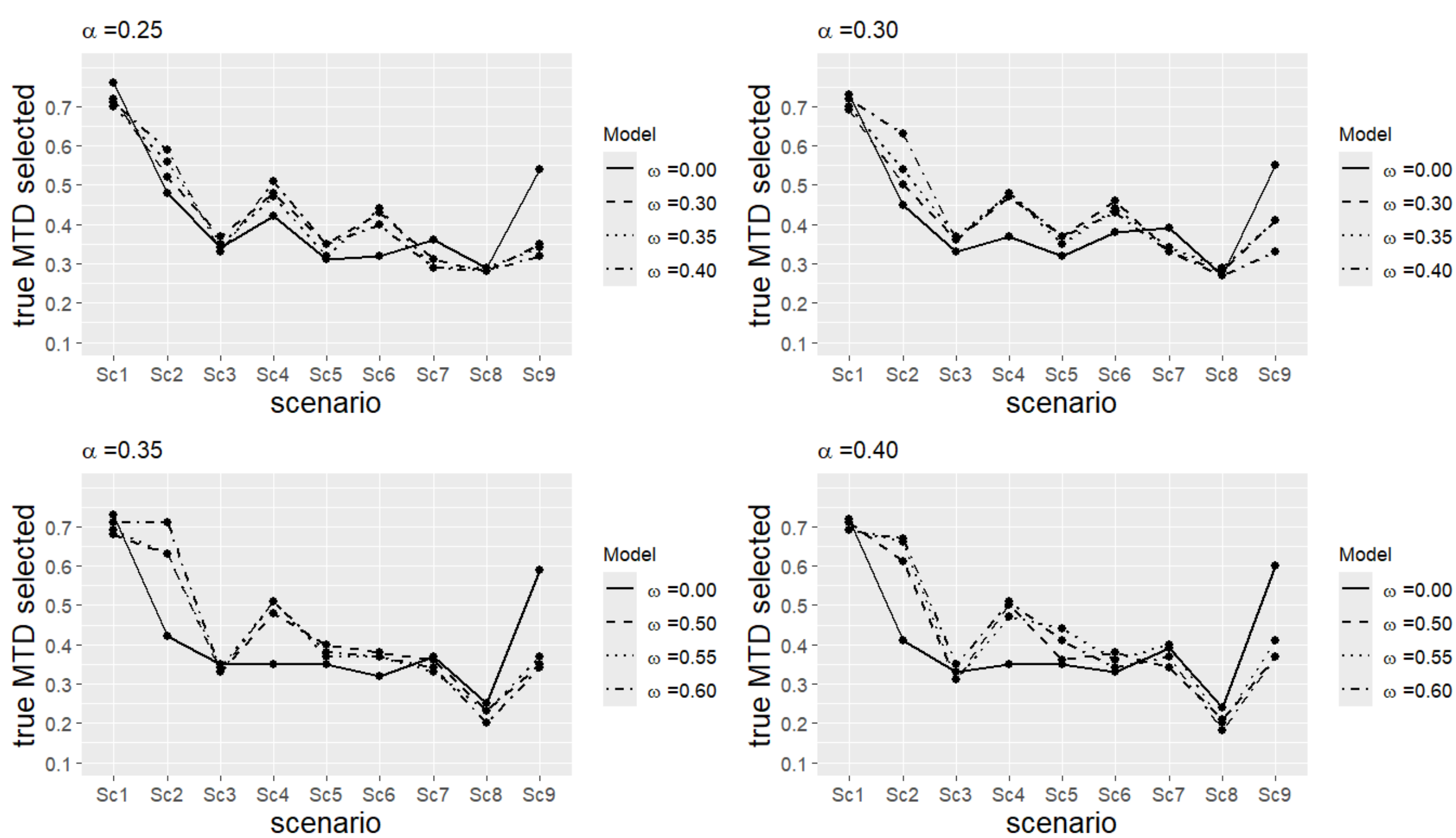

*Figure 6: Proportions of true MTD selected for different combinations of (α, ω), relation 1,* $d_{ref} = 22$.

From this calibration process, the parameter pair $(\alpha, \omega) = (0.35; 0.60)$ emerged as the most effective across both reference doses. Specifically, this combination provided the best trade-off between minimizing the proportion of overly toxic doses incorrectly selected as the MTD and maintaining a comparable rate of correct MTD identification. Tables 6 and 7 confirm that this $(\alpha, \omega)$ pair consistently yielded the lowest proportion of toxic selections while preserving accuracy in identifying the true MTD across multiple scenarios.

A similar calibration procedure was applied for Relation 2. For $d_{ref} = 70$, the optimal combination was identified as $(\alpha, \omega) = (0.25; 0.40)$, whereas for $d_{ref} = 22$, the best-performing pair was $(\alpha, \omega) = (0.35; 0.50)$. Detailed results for Relation 2, including the corresponding figures and tables, are available in the supplementary material.

These selected tuning parameter combinations will be used in subsequent sections to compare the BBLRM against other competing dose-escalation models.

To better understand how the parameters $\delta$ and $\omega$ influence the probability of toxicity, their burdening effect on the estimated DLT probability is shown below for the optimal combinations of $(\alpha, \omega)$ previously identified and applied within the BBLRM framework under Relation 1 and Relation 2, evaluated at the two reference doses, $d_{ref} = 22, 70$.

Specifically, Figures 7 and 8 illustrate the average posterior mean dose-toxicity curves across all simulated trials at the final cohort, stratified by scenario. Three models are compared: the standard BLRM without burdening (Zhang BLRM, black line), the BBLRM incorporating burdening under Relation 1 (red line), and under Relation 2 (green line).

A systematic trend is observed whereby the red curve (Relation 1) generally lies above the green curve (Relation 2), and both exceed the black curve across most dose levels and scenarios. This indicates that the incorporation of a burdening mechanism, particularly under Relation 1, yields increased posterior estimates of DLT probabilities.

Differences in the burdening effect are also observed between the two reference doses. When $d_{ref} = 22$, the burdening effect under BBLRM is more pronounced in later scenarios (Sc5–Sc9), whereas for $d_{ref} = 70$, the effect is more evident in early scenarios (Sc1–Sc3) and attenuates in later ones. These dynamics reflect model performance in terms of reducing the selection of excessively toxic doses as the MTD and in the identification rate of the true MTD, as detailed in the subsequent section.

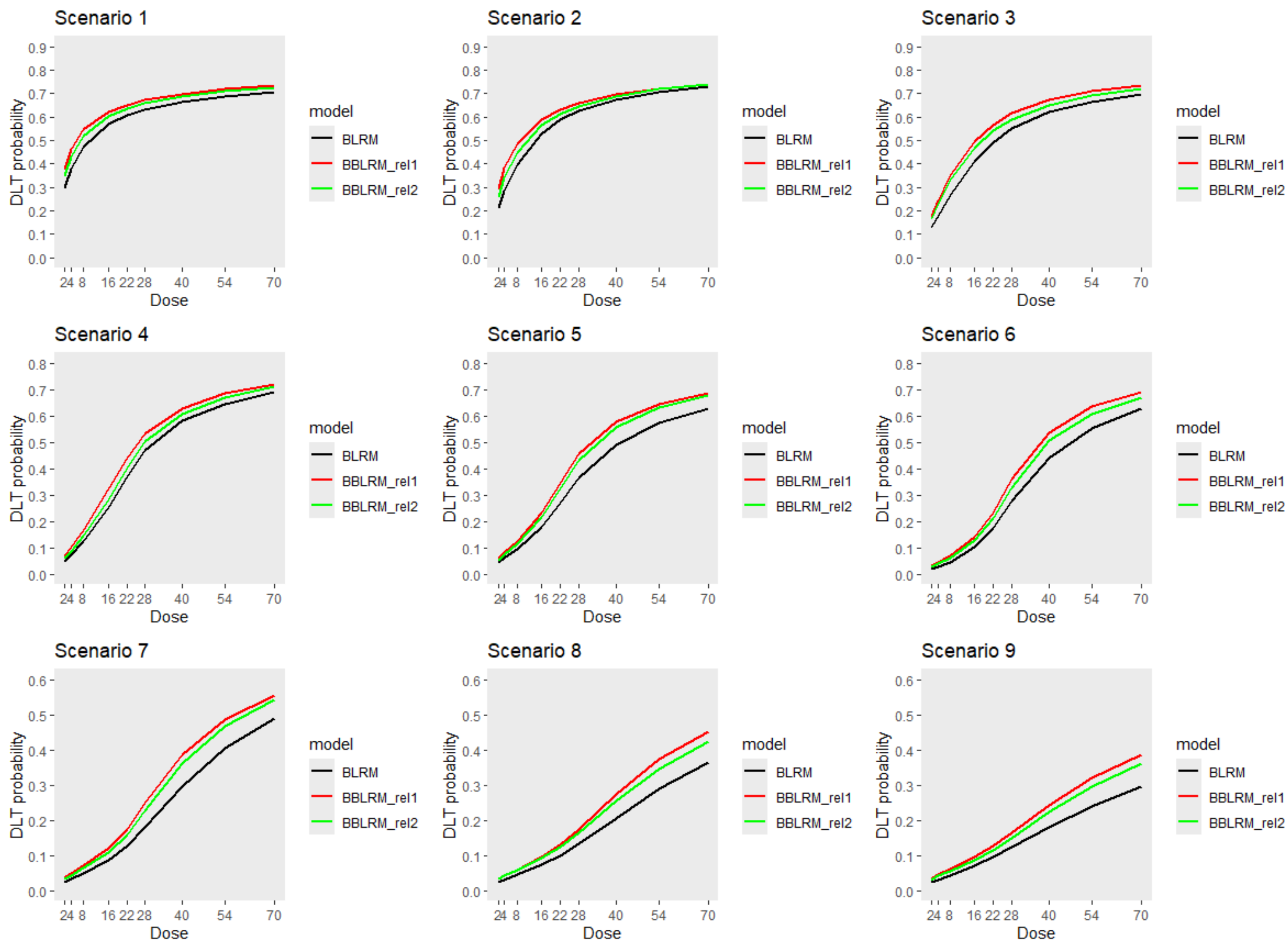

*Figure 7: Dose-toxicity curve for DLT probability related to Zhang BLRM (black line), BBLRM in Relation 1 (red line) and BBLRM in relation 2 (green line). The curves are obtained by averaging the posterior mean of DLT probability estimates of the last cohort of each simulated trail in each scenario,* $d_{ref} = 22$.

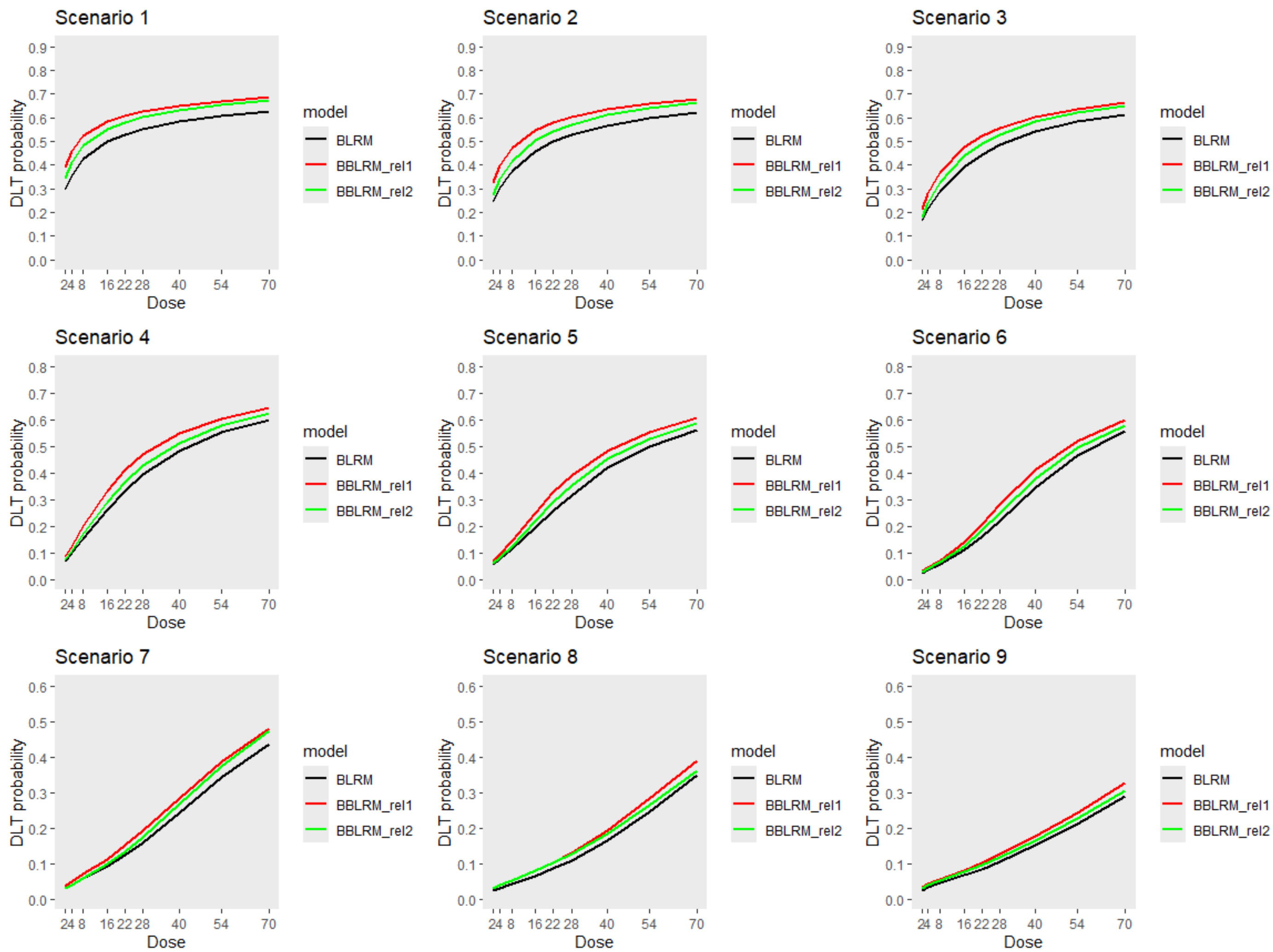

*Figure 8: Dose-toxicity curve for DLT probability related to Zhang BLRM (black line), BBLRM in Relation 1 (red line) and BBLRM in relation 2 (green line). The curves are obtained by averaging the posterior mean of DLT probability estimates of the last cohort of each simulated trail in each scenario,* $d_{ref} = 70$.

## 4. **Simulation results**

After calibrating Zhang's BLRM and BBLRM to determine the optimal tuning parameter values, as outlined in Section 3, we conducted additional simulations for two other competing models. In this section, the four models (i.e. BBLRM, Zhang's BLRM[22], two-stage CRM[25], and hCRM[24]) were compared based on the same operating characteristics considered during the calibration phase.

The figures and tables presented in this section summarise the results for Relation 1 and Relation 2, varying the reference dose of the BLRM model in the BBLRM, Zhang BLRM and hCRM. As stated earlier, the lines related to BBLRM refer to the optimal tuning parameter combinations obtained during the calibration phase for each assumed relationship.

Looking at the trends reported in Figure 9 and Figure 10, overall the BBLRM outperforms other models in minimizing the selection of toxic doses as MTD in most of the eight scenarios in both Relation 1 and 2 and with the two different reference doses except for scenarios Sc7 and Sc8 where the two-stage CRM performs better than the other models, especially when the $d_{ref} = 70$ (bottom panels of Figure 9).

Looking at the BBLRM performance, we can note that when $d_{ref} = 22$, the model reaches a higher reduction compared to when $d_{ref} = 70$. This behaviour, more visible in Relation 1, is not surprising considering how the two relations have been set. In the first one, we have that P(nDLTAE) increases as long as the P(DLT), thus incrementing the burdening effect of the $\delta$ parameter. On the other hand, in Relation 2, P(nDLTAE) increases up to the target DLT rate, then decreases up to the lowest plateau, thus reducing the burdening effect.

Regarding the two competing models, the two-stage CRM (Figure 9, dotted line) generally performs worse than BBLRM, except in Sc7 and Sc8. However, its performance is comparable to that of the Zhang BLRM model (Figure 9, green line). The hCRM model shows a distinct behaviour depending on the chosen reference dose. Specifically, when using either $d_{ref} = 22$ or $d_{ref} = 70$, the first four scenarios result in a high proportion of toxic doses being selected as the MTD. Starting from Scenario 5, a significant reduction in toxic dose allocation is observed when $d_{ref} = 22$ is used, while with $d_{ref} = 70$, the performance remains similar to the earlier scenarios (Figure 9, dotted line).

Focusing on the allocation of true MTDs, overall, the results show limited differences through the various models.

Focusing on Relation 1 (left panels of Figure 10), BBLRM with $d_{ref} = 22$ performs similarly to BBLRM with $d_{ref} = 70$ in the early scenarios (Sc1, Sc2, Sc3). It shows higher true MTD allocation in Sc4 and Sc5, but a lower proportion of true MTD selection in the later scenarios, particularly in Sc8 and Sc9.
Zhang BLRM displays more variability compared to BBLRM when using $d_{ref} = 22$, with better true MTD allocation in the later scenarios (Sc8 and Sc9). When using $d_{ref} = 70$, its performance aligns more closely with that of BBLRM.

Regarding Relation 2, there is no significant difference in correct MTD identification between BLRM and BBLRM.

Looking at the competing models, hCRM shows weaker performance in identifying the correct MTD in the early scenarios (Sc1–Sc3). However, it performs comparably to the other models in the middle scenarios, and in Sc9, it achieves the highest proportion of true MTD allocation (78%).

In contrast, while BBLRM, Zhang BLRM, and hCRM show more variable results across scenarios, the two-stage CRM delivers more consistent performance, outperforming the other models in Scenarios 3, 6, and 7 (Figure 10, dotted line) under both Relation 1 and Relation 2.

To provide an overall assessment of the models' performance across different scenarios, summary statistics (mean, median, minimum, and maximum) for the two operating characteristics are reported in Table 7.

Looking at the toxic doses selected as MTD (Table 7, left panel), BBLRM demonstrates the lowest values for the proportion of toxic dose allocation in Relation 1 and Relation 2, adopting both $d_{ref} = 22{,}70$ underscoring its robustness in preventing toxic dose selection.

The best results about proportion's reduction of toxic doses selected as MTD are related to Relation 1 for BBLRM with $d_{ref} = 22$, with mean (17%), median (16%), and minimum (9%) values, while when $d_{ref} = 70$, the reduction of toxic doses selection is slightly lower but still the best one compared to the other competing models. This may result from the specific relationship between P(DLT)and P(nDLTAE), where higher frequencies of nDLTAEs assumed in this relation impose a greater burden on DLT probabilities. Conversely, under Relation 2, where P(nDLTAE) decreases when P(DLT) $> 0.33$, the burden on DLT probability is reduced, slightly affecting BBLRM's ability to minimise toxic dose selection; however, the best reduction is still reached by the BBLRM model, especially with $d_{ref} = 22$. The two-stage CRM, instead, remains consistent across relationships, with stable summary statistics and performance similar to Zhang's BLRM. As previously mentioned, hCRM does not significantly reduce the proportion of toxic doses selected as MTD, exhibiting the highest values across all summary statistics.

Looking at the allocation of true MTD (Table 7, right panel), results reveal comparable overall performance for BBLRM, Zhang's BLRM, and the two-stage CRM. The first two models exhibit higher mean values (42-43%) across relationships but slightly lower median values, reflecting variability in performance across scenarios, especially in Relation 1. In contrast, the two-stage CRM achieves the highest median values (38–41%) with similar mean values, indicating more consistent performance across scenarios. The hCRM demonstrates comparable results but slightly underperforms BBLRM and Zhang's BLRM in terms of mean and the two-stage CRM in terms of median.

Based on the summary statistics reflecting the models' overall performance, the BBLRM delivers the best results, although in some individual scenarios, other models may perform better.

*Tables 7: Summary statistics related to proportions of toxic doses wrongly selected as MTD (left panel) and true MTD selected (right panel) for the different competing models in the different relationships.*

| Relation 1: toxic doses wrongly selected as MTD, $d_{ref} = 22$ | | | | |
|---|---|---|---|---|
| | Mean | Median | Min | Max |
| BBLRM | 0,17 | 0.16 | 0.09 | 0.29 |
| Zhang BLRM | 0.34 | 0.36 | 0.24 | 0.37 |
| 2 stage CRM | 37 % | 35 % | 23 % | 49 % |
| hCRM | 0.44 | 0.47 | 0.31 | 0.56 |
| **Relation 1: toxic doses wrongly selected as MTD, $d_{ref} = 70$** | | | | |
| | Mean | Median | Min | Max |
| BBLRM | 25 % | 26 % | 14 % | 36 % |
| Zhang BLRM | 36 % | 40 % | 26 % | 44 % |
| 2 stage CRM | 37 % | 35 % | 23 % | 49 % |
| hCRM | 53 % | 53 % | 47 % | 62 % |
| **Relation 2: toxic doses wrongly selected as MTD, $d_{ref} = 22$** | | | | |
| | Mean | Median | Min | Max |
| BBLRM | 0.23 | 0.24 | 0.14 | 0.28 |
| Zhang BLRM | 0.34 | 0.36 | 0.24 | 0.37 |
| 2 stage CRM | 38 % | 37 % | 22 % | 51 % |
| hCRM | 0.44 | 0.47 | 0.31 | 0.56 |
| **Relation 2: toxic doses wrongly selected as MTD, $d_{ref} = 70$** | | | | |
| | Mean | Median | Min | Max |
| BBLRM | 30 % | 32 % | 19 % | 41 % |
| Zhang BLRM | 36 % | 40 % | 26 % | 44 % |
| 2 stage CRM | 38 % | 37 % | 22 % | 51 % |
| hCRM | 53 % | 53 % | 47 % | 62 % |

| Relation 1: true MTD selected, $d_{ref} = 22$ | | | | |
|---|---|---|---|---|
| | Mean | Median | Min | Max |
| BBLRM | 0.43 | 0.37 | 0.20 | 0.71 |
| Zhang BLRM | 0.42 | 0.36 | 0.29 | 0.76 |
| 2 stage CRM | 39 % | 40 % | 27 % | 66 % |
| hCRM | 0.37 | 0.35 | 0.26 | 0.55 |
| **Relation 1: true MTD selected, $d_{ref} = 70$** | | | | |
| | Mean | Median | Min | Max |
| BBLRM | 43 % | 34 % | 24 % | 73 % |
| Zhang BLRM | 42 % | 34 % | 23 % | 72 % |
| 2 stage CRM | 39 % | 40 % | 27 % | 66 % |
| hCRM | 38 % | 32 % | 18 % | 78 % |
| **Relation 2: true MTD selected, $d_{ref} = 22$** | | | | |
| | Mean | Median | Min | Max |
| BBLRM | 0.42 | 0.41 | 0.25 | 0.72 |
| Zhang BLRM | 0.42 | 0.36 | 0.29 | 0.76 |
| 2 stage CRM | 39 % | 38 % | 27 % | 66 % |
| hCRM | 0.37 | 0.35 | 0.26 | 0.55 |
| **Relation 2: true MTD selected, $d_{ref} = 70$** | | | | |
| | Mean | Median | Min | Max |
| BBLRM | 42 % | 38 % | 24 % | 70 % |
| Zhang BLRM | 42 % | 34 % | 23 % | 72 % |
| 2 stage CRM | 39 % | 38 % | 27 % | 66 % |
| hCRM | 38 % | 32 % | 18 % | 78 % |

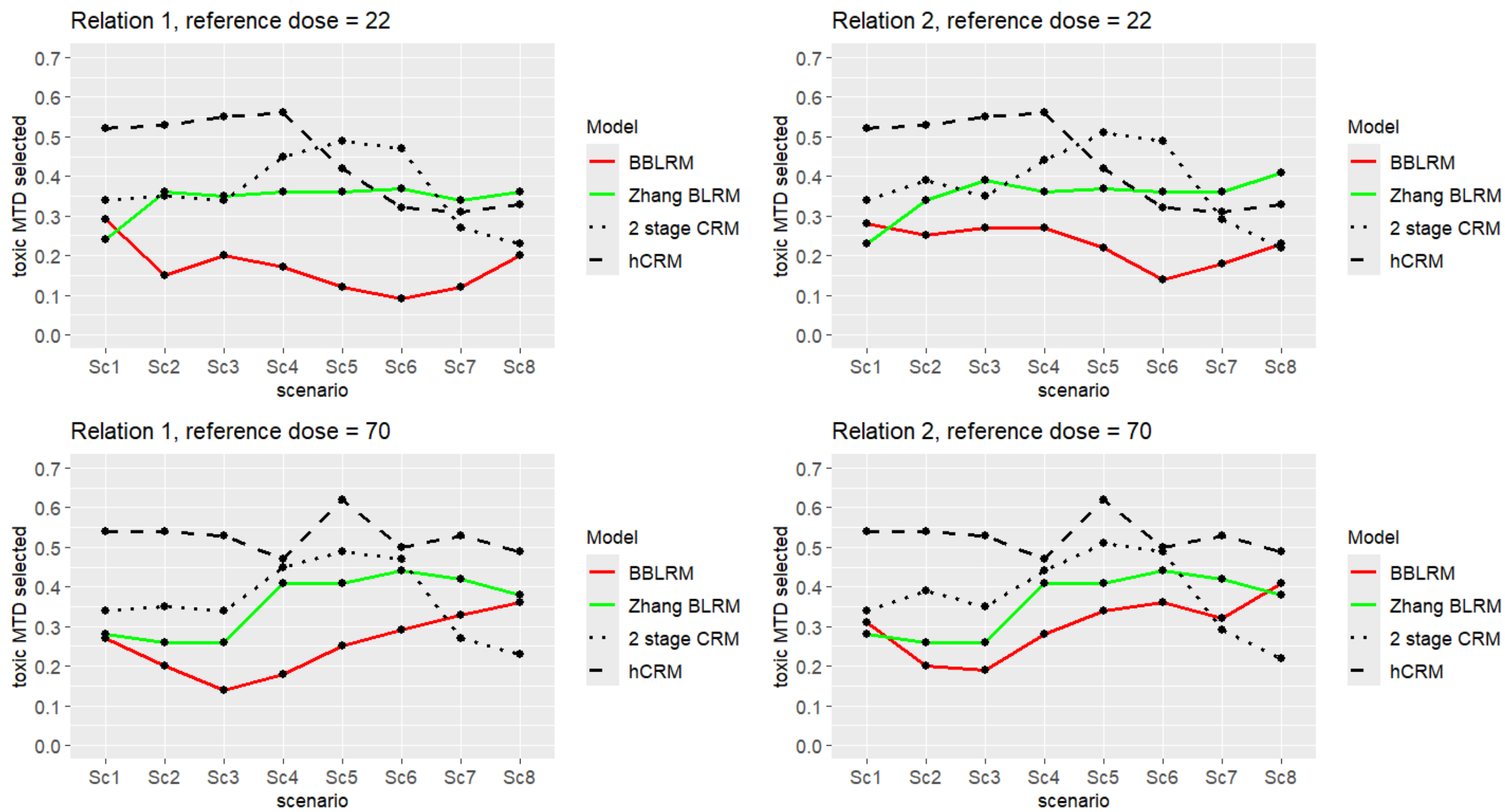

*Figure 9: Proportions of toxic doses wrongly selected as MTD for Relation 1, 2 and dose reference=22,70 for BBLRM, Zhang BLRM, 2-stage CRM and hCRM.*

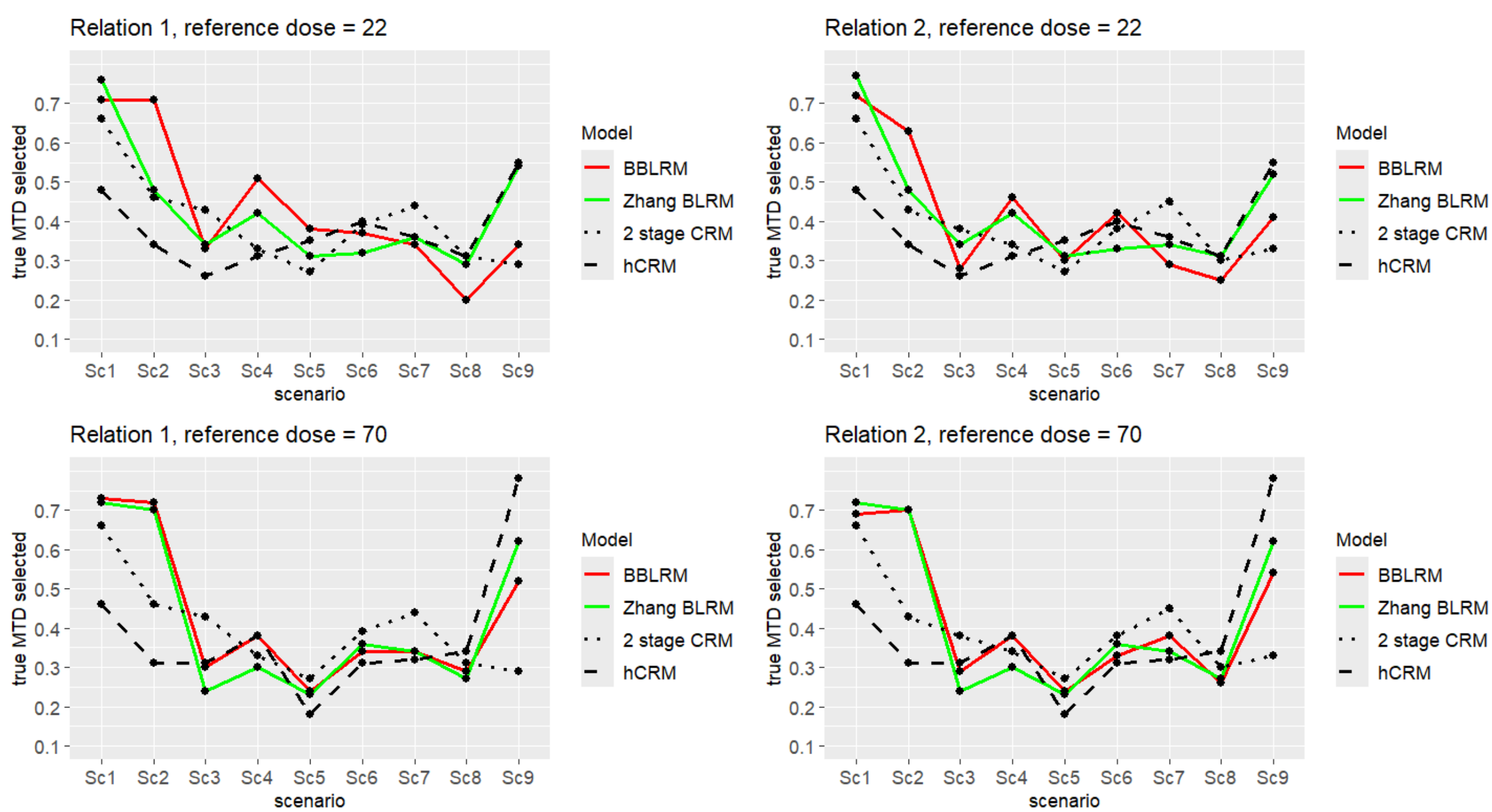

*Figure 10: Proportions of true MTD selected for Relation 1, 2 and dose reference=22,70 for BBLRM, Zhang BLRM, 2-stage CRM and hCRM..*

## 5. Discussion and conclusion

In this work, we aimed to enhance the performance of model-based approaches in phase I oncology trials by further developing the Bayesian logistic regression model (BLRM). Our primary objective was to improve patient safety by reducing the proportion of toxic doses selected as the maximum tolerated dose (MTD). To this end, we incorporated information about nDLTAEs that may be indicative of the subsequent occurrence of DLTs, adjusting the posterior probability of DLT occurrence based on the frequency of these related events. Clinician input was essential in identifying, a priori, which types of nDLTAEs should be considered, underscoring the importance of expert judgment in model development.

In the simulation study (Section 3), we explored two different approaches to generate DLT and nDLTAE outcomes that reflect the predictive behaviour typically observed.

Building on the recent BLRM model proposed by Zhang et al. in 2022[22], which served as our benchmark, we introduced a burden parameter that accounts for nDLTAEs observed within each cohort, resulting in the Bayesian burdened logistic regression model (BBLRM).

We evaluated the operating characteristics of the BBLRM through extensive simulations, considering a range of variables that may influence the model's performance. These included factors related to the trial context, such as the shape of the dose-toxicity curve, varying the MTD within the available dose levels, the relationship between DLT and nDLTAE probabilities, as well as aspects related to the model setup, including the calibration of tuning parameters and the selection of the reference dose.

The simulation study demonstrated an overall significant reduction in the rate of toxic doses allocated as MTD compared to competing models, such as Zhang's BLRM[22], the two-stage CRM[25], and hCRM[24]. Examining the specific scenarios, BBLRM is more effective in reducing the selection of toxic doses as MTD.

Two-stage CRM performs better when the MTD is located at the end of the dose vector. In general, Zhang's BLRM and hCRM do not outperform BBLRM across all scenarios. Notably, BBLRM maintained performance comparable to other models in correctly identifying the true MTD. Results vary slightly depending on the relationship between P(DLT) and P(nDLTAE) and on the reference dose adopted, reflecting different clinical contexts.

As outlined in the simulation settings, these results were achieved following a calibration phase to identify optimal values for the two BBLRM tuning parameters: $\alpha$, associated with Zhang et al.'s escalation rule[22], and $\omega$, linked to the burden parameter for estimating DLT probabilities. These parameters govern the model's conservatism from the escalation rules ($\alpha$) and model points of view ($\omega$), requiring careful calibration to balance between overly rapid and overly cautious dose escalation. Since there is no universal optimum for these parameters, trial-specific preliminary simulations are strongly recommended to identify the best combination of $\alpha$ and $\omega$, considering different factors subjective to the trial setting such as the P(DLT) and P(nDLTAE) relationship, dose reference, dose vector, number of cohorts and cohort sizes.

In conclusion, we introduce the BBLRM as an extension of BLRM, designed to enhance safety in phase I oncology trials by reducing the probability of selecting toxic doses as MTD without compromising the identification of the true MTD. By actively involving clinicians in defining nDLTAEs before starting the trial, this approach fosters a more collaborative and precise dose-escalation process. We hope this development will encourage the broader adoption of model-based methods in clinical practice, enabling a safer and more efficient determination of MTDs in oncology trials.

**Declaration of conflicting interest**

The author(s) declared no potential conflicts of interest with respect to the research, authorship, and/or publication of this article.

**Funding statement**

The author(s) received no financial support for the research, authorship, and/or publication of this article.

**Data availability**

An R code example to reproduce the results shown in the "Simulations" section is available in the Supplementary Material.